# Tap Tips: Lightweight Discovery of Touchscreen Targets

Paul M. Aoki     Amy Hurst     Allison Woodruff

Xerox Palo Alto Research Center
3333 Coyote Hill Road
Palo Alto, CA 94304-1314 USA

**ABSTRACT**
We describe tap tips, a technique for providing touch-screen target location hints. Tap tips are lightweight in that they are non-modal, appear only when needed, require a minimal number of user gestures, and do not add to the standard touchscreen gesture vocabulary. We discuss our implementation of tap tips in an electronic guidebook system and some usability test results.

## INTRODUCTION

Providing *usage tips* is not always straightforward for devices, such as personal digital assistants (PDAs), that are based on a pen user interface (PUI). Usage tips help the user understand the location and purpose of interface components and can take the form of cursor changes, status bar messages, or transient pop-up windows ("tooltips"). Typical GUI-based systems provide usage tips when the user performs a "mouse-over" gesture. Since PUIs do not have a natural analogue of mouse-over,[1] designers must find new approaches for triggering tips.

The PUI usage tip problem becomes particularly acute in the context of *imagemaps* – images with *targets*, or "hot" regions, that cause some action to be performed when the user selects them. Imagemaps are common in Web pages and other graphical hypertext applications. Location-revealing usage tips are important for imagemaps because the locations of targets may not be obvious to users.

In this paper, we describe *tap tips*, a novel mechanism for providing imagemap usage tips for touchscreen interfaces. Tap tips are lightweight for the end-user – they are non-modal, appear only when needed, require a minimal number of user gestures, and do not expand the standard PUI gesture vocabulary – and require only standard touchscreen hardware. We have implemented tap tips in a PDA-based electronic guidebook system and have received promising feedback from multiple rounds of usability tests.

---

[1] GUI mouse-over involves moving the cursor over a region of the screen and then momentarily holding it still. Standard PDA hardware supports only a binary notion of contact with the screen (the stylus is either in contact with a pressure-sensitive surface or it is not), and therefore standard PUIs do not have a "hover" gesture equivalent to GUI mouse-over.

## KNOWN APPROACHES ARE UNSATISFACTORY

Current solutions, as well as solutions based on obvious extensions to existing approaches, all have significant disadvantages. We now discuss the non-modal and modal alternatives in turn, excluding those that require additional hardware (e.g., magnetic proximity sensors).

The obvious non-modal alternatives are to (1) not show tips at all, or (2) show tips at all times. A complete lack of tips reduces the user to a frustrating, trial-and-error experience. Continuously-present tips effectively change the imagemap content. For example, some applications identify targets by making them look like 3D button controls; others use outlines. Such approaches are visually distracting, aesthetically unappealing, and incur a pixel overhead that can be prohibitive for mobile devices. They may also suggest a "checklist" approach to imagemap usage that draws attention away from the actual task.

The main modal alternatives are (1) a straightforward "tip mode" using standard user gestures, or (2) transient tips invoked (and dismissed) by special user gestures. Both have significant overheads. Activating a tip mode requires the user to remember and then execute two additional gestures (mode entry and mode exit). The most common transient tip methods are variations on "slide to see, lift to do" [2,3,5]. In this approach, tips appear when the user presses the stylus over a given screen region for a certain amount of time; the target is selected when the stylus is *lifted*. The user must learn a novel, task-specific gesture – one that is relatively unintuitive because it violates the pen metaphor.

## TAP TIPS

The previous discussion illustrates that an effective locational tip mechanism must provide information to the user while simultaneously minimizing the number of gestures that must be performed, the number of gestures the user is required to learn and remember, and the amount of time that tips are present.

A non-modal method with the following characteristics avoids many problems described above. When the user performs a triggering gesture, the system provides feedback that *indicates the location of the targets*. The indication *has finite duration,* ceasing without an explicit, separate gesture by the user. (Of course, the system may interrupt the feedback in response to subsequent user actions.)

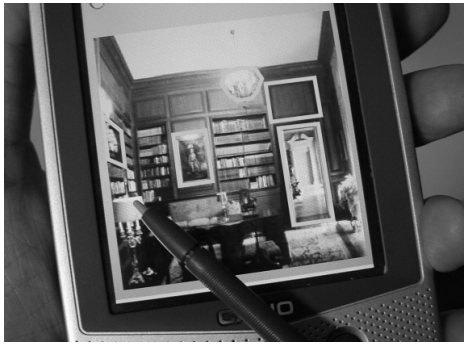

**Figure 1**. Transient target outlines.

We define tap tips as transient locational hints that are triggered by failure to select a target – a target "miss." This approach has several benefits. First, the location hints are generally provided in the exact situation where they are needed (i.e., failure demonstrates that the user does not know the precise location of the targets). Second, no novel gestures must be learned and no additional gestures are required – the location hints are provided as a side-effect of a gesture that was already made (i.e., the tap that missed).

**An Example**
As a concrete example, we describe an implementation of these concepts in an electronic guidebook for historic houses running on a Casio Cassiopeia E-105 color PDA. A location-based context-aware interface is infeasible in this domain for reasons described elsewhere [1]. Therefore, the system presents the visitor with one of a collection of imagemaps. Each imagemap is a photograph of one wall in a room. The visitor manually navigates to the desired wall. When the visitor taps on a target, the guidebook displays a text description or plays an audio clip.

Many of the objects visible in a given imagemap have descriptions. However, not all do, and not all of the described objects are as "obvious" as they would be in (e.g.) a museum; Figure 1 shows one imagemap in which a wood panel and a doorway have associated descriptions. Providing feedback about the imagemap targets is therefore critical in preventing visitor frustration.

The guidebook displays outlines around each imagemap target, triggered when the user taps on the imagemap but does not "hit" a target. The outlines appear, gradually fade and finally disappear. The outlines are drawn in colors chosen for their visual "popout" effectiveness [4] and are visible for under 2 seconds. We visually distinguish targets that have been previously selected from those that have not.

**USABILITY RESULTS AND DISCUSSION**
We conducted three rounds of usability testing. Each participant performed a self-guided tour of one or two rooms using our prototype. The first two rounds involved semi-structured post-tour interviews with a total of thirteen subjects. The third round involved fourteen subjects (most of whom had a non-technical background and were not active PDA users) ranging in age from 7 to "over 60." We recorded the participants on audio and video, logged their user interface gestures on the PDA, and conducted semi-structured post-tour interviews.

The interviews provided some encouraging high-level feedback. All of the participants in all three rounds were able to use the prototype to obtain descriptions of the objects around them. None expressed confusion about the target discovery interface, though one participant initially had difficulty understanding the imagemap concept.

Analysis of the device logs provided additional positive usability evidence. As mentioned, many targets were not necessarily visually obvious. However, all participants were able to use the interface to discover multiple targets and select them without requiring additional tips to refresh their memory. Several selected six targets in a row.

Observations suggest that tap tips are effective when the underlying imagemap approach works (reasonably) well. For example, this technique works well when enough of the targets are sufficiently "obvious" that tips are needed occasionally rather than continuously. Obvious objects include large or "important-looking" objects (e.g., paintings) that have visual and/or semantic popout. Tap tips were successful in guiding users to less obvious targets as well as providing feedback when plausible targets did not have descriptions. An example where imagemaps were less effective was when targets were numerous, encouraging a checklist-like behavior. In this situation, some tap tip users alternated between viewing the tips and selecting a target (even when only one object remained unselected!). A small number of users who exhibited this behavior requested a tip mode.

In short, our results indicate that tap tips work well for their designed purpose as an assistive mechanism.